\documentstyle{lamuphys}
\input{psfig}
\makeatletter
\let\chapter\hid@chapter
\makeatother
\begin{document}
\pagenumbering{arabic}

\title{Cosmological Tests from the New Surveys }

\author{~~Ofer Lahav}

\institute{Institute of Astronomy, Madingley Road,
Cambridge CB3 0HA, UK } 

\maketitle

\begin{abstract}
  We review cosmological inference from galaxy surveys at low and
  high redshifts, with emphasis on new Southern sky surveys.  We focus
  on several issues: (i) The importance of understanding selection
  effects in catalogues and matching Northern and Southern surveys;
  (ii) The 2dF galaxy redshift survey of 250,000 galaxies (iii) The
  proposed 6dF redshift and peculiar velocity survey of near-infrared
  galaxies (iv) Radio sources and the X-Ray Background as useful
  probes of the density fluctuations on large scales, and (v) How
  to combine large scale structure and Cosmic Microwave Background
  measurements to estimate cosmological parameters.
\footnote{Invited talk, to appear in the Proceedings of the ESO/ATNF Workshop
{\it Looking Deep in the Southern Sky}, 10-12 December 1997, Sydney, Australia,
Eds. R. Morganti  and W. Couch}
\end{abstract}

\section{Introduction}

It is believed by most cosmologists that 
on the very large scales the universe 
is an isotropic and homogeneous system. 
However, on scales much smaller than the horizon the distribution 
of luminous matter is clumpy.
Galaxy surveys in the last decade have provided a major tool for
cosmographical and cosmological studies.  In particular, surveys such
as CfA, SSRS, IRAS, APM and Las Campanas have yielded useful information on
local structure and on the density parameter  $\Omega$ from
redshift distortion and from comparison with the peculiar velocity
field.  Together with measurements of the Cosmic Microwave Background
(CMB) radiation and gravitational lensing the redshift surveys provide
major probes of the world's geometry and the dark matter.

In spite of the rapid progress
there are two gaps in our current understanding of 
the density fluctuations as a function of scale:
(i) It is still unclear how to relate the distributions of light 
    and mass, in particular how to match 
    the clustering of galaxies with the CMB fluctuations,
(ii) Currently little is known about fluctuations 
on  intermediate scales 
between these of local galaxy surveys ($\sim 100 h^{-1} $ Mpc)
and the scales probed by COBE ($\sim 1000 h^{-1} $ Mpc). 

Another unresolved issue is the value of the density parameter
$\Omega$.  Putting together different cosmological observations, the
derived values seem to be inconsistent with each other.  
Taking into account moderate
biasing, the redshift and peculiar velocity data on large scales yield
$\Omega \approx 0.3 -1.5$, with a trend towards the popular 
value $\sim 1$
(e.g. Dekel 1994; Strauss \& Willick 1995 for summary of results).
On the other hand, the high fraction of baryons in clusters, combined
with the baryon density from Big Big Nucleosynthesis suggests $\Omega
\approx 0.2$ (White et al. 1993).  Moreover, an $\Omega=1$ universe is
also in conflict with a high value of the
Hubble constant ($H_0 \approx 70-80$ km/sec/Mpc),
as in this model the universe turns out to be younger
than globular clusters.  A   way out of these problems 
was suggested by 
adding a positive cosmological constant, such that $\Omega + \lambda
=1$, to satisfy inflation.  Two recent observations 
constrain 
$\lambda $ : the observed frequency of lensed quasars is too small,
yielding an upper limit $\lambda <0.65$ 
(e.g. Kochanek 1996),
and the
magnitude-redshift relation for Supernovae type Ia 
(e.g. Perlmutter et al. 1998).
The next decade will
see several CMB experiments (e.g. Planck, MAP, VSA) which promise to
determine (in a model-dependent way) the cosmological parameters to
within  a few percent.
We shall focus here on several issues related to clustering and
cosmological parameters from new surveys.

\section{From `Biased Surveys' to the `Real Universe'}

Figure 1 shows a compilation of the current and future surveys,
indicating for each survey its effective volume (in terms of 
its solid angle and median redshift) and the number of galaxies
with measured redshift.

\begin{figure}
\protect\centerline{
\psfig{figure=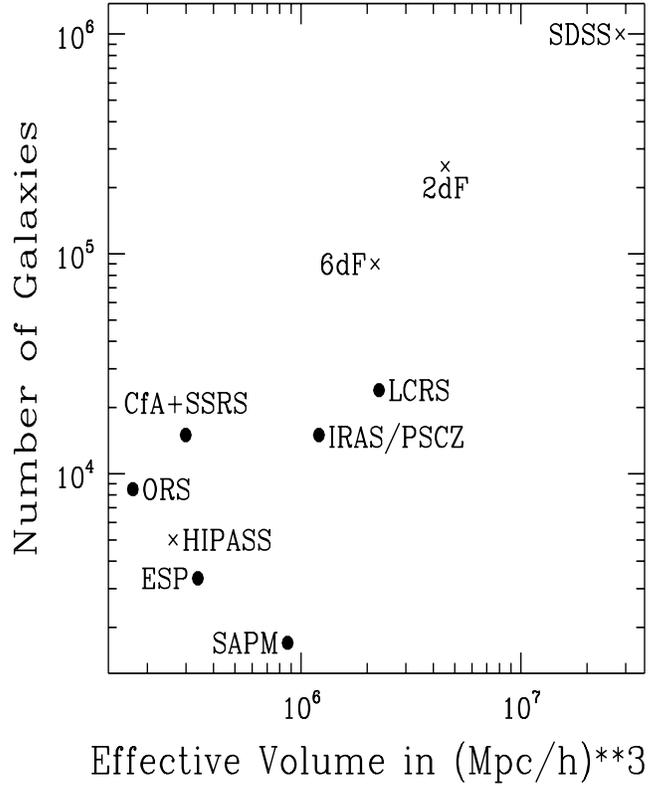,height=5truein,width=4truein}}
\caption[]{The effective volume and
number of galaxies of
completed reshift surveys (solid circles)
and surveys in preparation (crosses).
The effective volume is defined here 
as $\omega R_{med}^3/3$, where $\omega$ is the solid angle of a
survey, and $R_{med}$ its median comoving depth. 
Optically selected surveys include the Center for Astrophysics 
(CfA) survey, the Southern Sky Redshift
Survey (SSRS), the Optical 
Redshift Survey(ORS), the ESO Slice
Project(ESP), the Stromlo-APM (SAPM) and the Las Campans Rdshift
Survey (LCRS), the 2-degree Field (2dF) galaxy survey, 
and the Sloan Digital Sky
Survey (SDSS). The IRAS Point Source Catalogue (PSCZ) is selected in
the infrared ($60 \mu$), 
while 6dF/DENIS/2MASS is selected in the near infrared $(\sim 2 \mu)$.
HIPASS is the Parkes multi-beam survey in 21 cm.
(A compilation by S. Maddox and O. Lahav).  } 
\end{figure}

It is important to emphasize that each survey is selected 
by different criteria (e.g. wavelength and flux), hence
any survey is biased ! 
One should pay attention 
to the following aspects in  analyzing surveys:

\hangindent \parindent \hangafter 1 
$\bullet$
Source Detection 
(e.g. 3-$\sigma$ selection, wavelet filtering, combining 
radio multi-components)

\hangindent \parindent \hangafter 1 
$\bullet$
Source Classification (e.g. star/galaxy separation - 
by eye or by Artificial Neural Networks)
and multi-wavelength identification

\hangindent \parindent \hangafter 1 
$\bullet$
Incomplete Sky Coverage (e.g. the Zone of Avoidance)

\hangindent \parindent \hangafter 1 
$\bullet$
Matching Northern and Southern catalogues
(e.g. the optical UGC/ESO, the radio 87GB/PMN)

\hangindent \parindent \hangafter 1 
$\bullet$
Poisson shot-noise, due to the finite number of galaxies

\hangindent \parindent \hangafter 1 
$\bullet$
Redshift distortion

\hangindent \parindent \hangafter 1 
$\bullet$
Biasing of particular tracer relative to the underlying mass distribution  

\section {New Surveys}
Existing optical and IRAS redshift surveys contain 10,000-20,000
galaxies.  
The Parkes multi-beam survey in 21cm (HIPASS) 
will detect about  5,000 galaxies
in the Southern hemisphere (see Steveley-Smith in this volume).
Another major step forward using multifibre technology will allow
in the near future to produce redshift surveys of millions of
galaxies.  In particular, there are two major surveys on the horizon.
The American-Japanese 
 Sloan Digital Sky Survey (SDSS) will yield images in 5
colours for 50 million galaxies, 
and redshifts for about 1 million galaxies over a quarter of the
sky (Gunn and Weinberg 1995). It will
be carried out using a dedicated 2.5m telescope in New Mexico.  The
median redshift of the survey is $z \sim 0.1$.
A complementary Anglo-Australian survey, the 2 degree Field (2dF), is
described below (see also Boyle \& Colless on both the galaxy and
quasar 2dF surveys in this volume).

\subsection {2dF}

The 2dF galaxy survey 
will produce redshifts for 250,000 galaxies brighter than $b_J =19.5^m$
(with median redshift of $z \sim 0.1$), selected from the APM catalogue.  The
survey will utilize a new 400-fibre system on the 4m AAT, covering
$\sim 1,700$ sq deg of the sky.  
About 8,000 redshifts have been measured so far (March 1998).
A deeper
extension down to $R=21$ for 10,000 galaxies is also planned for the 2dF 
survey.

\hangindent \parindent \hangafter 1
$\bullet$ Accurate measurements of the power spectrum of galaxy
clustering on scales $ > 30 h^{-1}$ Mpc allowing a direct
comparison with CMB  anisotropy measurements
such as the recently approved  NASA MAP and ESA Planck Surveyor
satellites. The power-spectrum derived from the projected APM galaxies
(see Figure 2) gives an idea about the scales probed by the 2dF
redshift survey. 

\hangindent \parindent \hangafter 1
$\bullet$ Measurement of the distortion of the clustering pattern in
redshift space providing constraints on the cosmological density
parameter $\Omega$ and the spatial distribution of dark matter.

\hangindent \parindent \hangafter 1
$\bullet$ Determination of variations in the spatial and velocity
distributions of galaxies as a function of luminosity, spectral type and
star-formation history, providing important constraints on
models of galaxy formation.

\hangindent \parindent \hangafter 1
$\bullet$ Investigations of the morphology of galaxy clustering and
the statistical properties of the fluctuations, {\it e.g.} whether the
initial fluctuations are Gaussian as predicted by inflationary models
of the early universe.

\hangindent \parindent \hangafter 1
$\bullet$ A study of clusters and groups of galaxies in the redshift
survey, in particular the measurement of infall in clusters and
dynamical estimates of cluster masses at large radii.

\hangindent \parindent \hangafter 1
$\bullet$ Application of novel techniques 
(e.g. Principal Component Analysis 
and Artificial Neural Networks; Folkes, Lahav \& Maddox 1996) 
to classify the uniform
sample of $250,000$ spectra obtained in the survey, thereby obtaining
a comprehensive inventory of galaxy types as a function of spatial
position within the survey.

For more details on the 2dF galaxy survey see

{\tt http://msowww.anu.edu.au/colless/2dF/}

\subsection {6dF}

  It is has recently been proposed by the Anglo-Australian Observatory
  to automate the FLAIR multi-fibre facility at the 1.2m
  Schmidt telescope in
  Siding Spring Australia. The main purpose of the upgrading is to
  measure redshifts to $\sim 120,000$ galaxies principally selected in
  the Near InfraRed from the DENIS survey, and to measure internal
  motions (hence distance indicators and peculiar velocities) to $\sim
  18,000$ galaxies.  The unique feature of this survey is the ability
  to probe {\it mass} on both galactic and cosmological scales.

  The European DENIS project (DEep Near-Infrared Southern Sky
  Survey) has begun in 1995 to image the entire Southern sky in
  three bands in the Near InfraRed (NIR): $I$(0.8 micron),
  $J$(1.25 micron) and $K_s$ (2.15 micron).  NIR light from
  galaxies is dominated by the old stellar population, hence is
  more directly related to the underlying mass than surveys in
  the optical or far-infrared. Moreover, the NIR light is little
  affected by Galactic extinction, making it ideal to probe
  galaxies through the Zone of Avoidance. Another large survey at
  2 microns (2MASS) is carried out by a US team.
  In the case of DENIS the main
  purpose of the survey is to measure redshift to 90,000 NIR-selected
  galaxies brighter than $J=13.7$ 
  (with median redshift ${\bar z}  \approx0.04$)
  plus perhaps 30,000 additional
  galaxies over 18000 deg$^2$ in about 160 spectroscopic nights.  Even
  more exciting than this big redshift survey is the possibility to
  measure velocity dispersions for elliptical galaxies and rotation velocities
  for spiral galaxies.  This will produce a uniform set of $\sim
  18000$ galaxies with distance indicators (via the Tully-Fisher
  relation), and hence peculiar velocities.  Currently the most recent
  sample of peculiar velocities (Mark III) includes only $\sim 3000$
  galaxies taken from various subsets.  The combination of the
  redshift and peculiar velocity surveys over an entire hemisphere
  will allow us to reconstruct 
   maps of NIR galaxy and mass distributions, 
  e.g. by the Potent (Dekel 1994)
  and Wiener (e.g. Webster et al. 1997) methods,
  to estimate the density
  parameter $\Omega$ and to characterize biasing. 
  Assuming that it will take 2 years to build 6dF, 
  the entire redshift survey can be finished by 2001
  and the peculiar velocity survey by 2003.
  Detailed   proposals can be found on
  {\tt http://msowww.anu.edu.au/colless/6dF/}

\bigskip

 \section {Probes at High Redshift}

 The big new surveys 
 (SDSS, 2dF) 
 will only probe a median redshift ${\bar z} \sim 0.1$.
 It is still crucial to probe the density fluctuations at higher $z$,
 and  to fill in the gap between
 scales probed by previous local galaxy surveys and the scales 
 probed by COBE and other CMB experiments.
 Here we discuss the X-ray Background (XRB) and radio sources 
 as probes of the density fluctuations at median redshift $ {\bar z} \sim 1$.
 Other possible high-redshift traces are quasars and clusters of galaxies.

 \subsection{Radio Sources}

Radio sources in surveys have typical median redshift
${\bar z} \sim 1$, and hence are useful probes of clustering at high
redshift. 
Unfortunately, it is difficult to obtain distance information from
these surveys: the radio luminosity function is very broad, and it is
difficult to measure optical redshifts of distant radio sources.
Earlier studies
claimed that  the distribution of radio sources supports the 
'Cosmological Principle'.
However, the redshift distribution of radio sources is now 
better understood, and it is 
clear that the wide range in intrinsic luminosities of radio sources
would dilute any clustering when projected on the sky.  
In fact, recent analyses  of
new deep radio surveys (e.g. FIRST)
suggest that radio sources are indeed clustered at least as strongly
as local optical
galaxies 
(e.g. Cress et al. 1996; Magliocchetti et al. 1998 and in this volume).
Nevertheless, on the very large scales the distribution of radio sources
seems nearly isotropic. 
Comparison of the measured quadrupole in a radio sample 
in the Green Bank and Parkes-MIT-NRAO
4.85 GHz surveys 
to the theoretically predicted ones (Baleisis et al. 1998)
offers a crude estimate of the fluctuations on scales $ \lambda \sim 600
h^{-1}$ Mpc.  The derived amplitudes are shown in Figure 2 for the two 
assumed Cold Dark Matter (CDM) models.
Given the problems of catalogue matching and shot-noise, these points should be
interpreted at best as `upper limits', not as detections.  
A new Southern radio survey, SUMSS, is described by Sadler in this volume.

 \subsection {XRB}

 Although discovered in 1962, the origin of
 the X-ray Background (XRB) is still unknown,  
 but is likely
 to be due to sources at high redshift 
 (for review see Boldt 1987; Fabian \& Barcons 1992).
 Here we shall not attempt to speculate on the nature of the XRB sources.
 Instead, we {\it utilise} the XRB as a probe of the density fluctuations at
 high redshift.  The XRB sources are probably
 located at redshift $z < 5$, making them convenient tracers of the mass
 distribution on scales intermediate between those in the CMB as probed
 by COBE, and those probed by optical and IRAS redshift
 surveys (see Figure 2).

The interpretation of the results depends somewhat on the nature of
the X-ray sources and their evolution.  The rms dipole and higher
moments of spherical harmonics can be predicted (Lahav et al. 
1997) in the
framework of growth of structure by gravitational instability from
initial density fluctuations.
By comparing
the predicted multipoles to those observed by HEAO1 
(Treyer et al. 1998)
it is possible to estimate the amplitude of fluctuations for an
assumed shape of the density fluctuations 
(e.g. CDM model).  
Figure 2 shows the amplitude of fluctuations derived at the 
effective scale $\lambda \sim 600 h^{-1}$ Mpc probed by the XRB. 
One can  use this estimate to derive the  fractal
dimension $D_2$ of the universe on large scales.
Although the fractal
dimension of the galaxy distribution on scales $< 20 h^{-1}$ Mpc is $D_2
\approx 1.2-2.2$, the fluctuations in the X-ray Background and in the
Cosmic Microwave Background are consistent with $D_2=3$ to within
$10^{-4}$  on the very
large scales (Wu et al. 1998).

\begin{figure}
\protect\centerline{
\psfig{figure=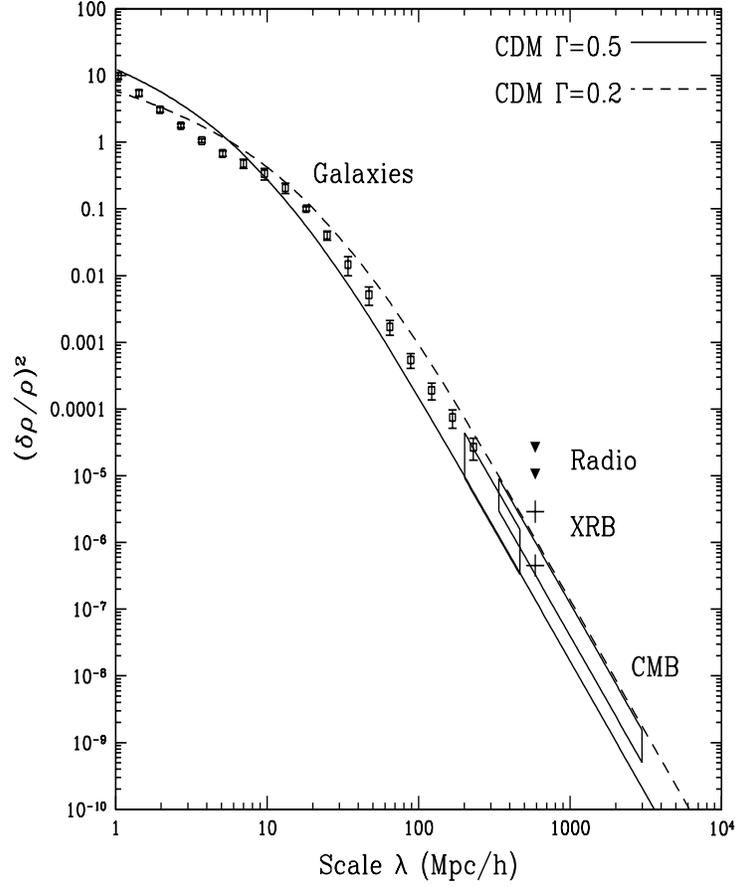,height=5truein,width=4truein}}
\caption[]{
  A compilation of density fluctuations on different scales from
  various observations: a galaxy survey, deep radio surveys, the X-ray
  Background and Cosmic Microwave Background experiments. The
  measurements are compared with two popular Cold Dark Matter models.
  The Figure shows mean-square density fluctuations $({ {\delta \rho}
    \over \rho })^2 \propto k^3 P(k)$, where $k=1/\lambda$ is the
  wavenumber and $P(k)$ is the power-spectrum of fluctuations.  The
  solid and dashed lines correspond to the standard Cold Dark Matter
  power-spectrum (with shape parameter $\Gamma = 0.5$) and a
  `low-density' CDM power-spectrum (with $\Gamma=0.2$), respectively.
  Both models are normalized such that the rms mass fluctuation at 
  $8 h^{-1}$ Mpc
  spheres is $\sigma_{8}=1$ (at $k \sim 0.15$).  The open squares at
  small scales are estimates of the power-spectrum from 3D inversion
  of the angular APM galaxy catalogue (Baugh \& Efstathiou 1994).
  The elongated 'boxes' at large scales represent the COBE
  4-yr (on the right) and
  Tenerife (on the left) CMB measurements.  The
  COBE 'box' corresponds to a quadrupole Q=18.0 $\mu K$ for a
  Harrison-Zeldovich mass power-spectrum, via the Sachs-Wolfe effect,
  or $\sigma_{8} = 1.4$ for a standard CDM model (Gawiser \& Silk
  1998).  The solid triangles represent constraints from the
  distribution of radio sources brighter than 70 mJy from the PMN and
  87GB samples.  This quadrupole  (Baleisis et al. 1998)
  probes fluctuations on scale $\lambda_*^{-1} \sim 600 h^{-1}$ Mpc.
  The top and bottom solid triangles are upper limits of the amplitude
  of the power-spectrum at $\lambda_*$, assuming CDM power-spectra
  with shape parameters $\Gamma=0.2$ and $0.5$ respectively, and an
  Einstein-de Sitter universe.
  The crosses represent constraints from the XRB HEAO1 quadrupole
  (Lahav et al. 1997; Treyer et al. 1998).  Assuming evolution,
  clustering and epoch-dependent biasing prescriptions, this XRB
  quadrupole measurement probes fluctuations on scale $\lambda_*^{-1}
  \sim 600 h^{-1}$ Mpc, very similar to the scale probed by the radio
  sources.  The top and bottom crosses are estimates of the amplitude
  of the power-spectrum at $\lambda_*$, assuming CDM power-spectra
  with shape parameters $\Gamma=0.2$ and $0.5$ respectively, and an
  Einstein-de Sitter universe.  The fractional error on the XRB
  amplitudes (due to the shot-noise of the X-ray sources) is about 30
  \%. (A compilation from Wu, Lahav \&Rees 1998).  }
\end{figure}

 \section {Joint Parameter Estimation LSS/CMB}

Observations of anisotropies in the Cosmic Microwave Background (CMB)
provide one of the key constraints on cosmological models and a
significant quantity of experimental data already exists 
(e.g. Gawiser \& Silk 1998 
and Lineweaver in this volume).

Galaxy redshift surveys, mapping large scale structure (LSS), provide
another cosmologically important set of observations.  The clustering
of galaxies in redshift-space is systematically different from that in
real-space (Kaiser 1987, Hamilton 1997 for review).
 The mapping between the two is a
function of the underlying mass distribution, in which the galaxies
are not only mass tracers, but also velocity test particles.
Estimates derived separately from each of the CMB and LSS  data sets have
problems with parameter degeneracy. 
Webster et al. (1998) combined results from a 
a range of CMB experiments, with a likelihood analysis of the IRAS
1.2Jy survey, performed in spherical harmonics.  
Their method expresses the effects of the
underlying mass distribution on both the CMB potential fluctuations
and the IRAS redshift distortion. This breaks the degeneracy inherent
in an isolated analysis of either data set, and places tight
constraints on several cosmological parameters. 

The family of CDM models analysed corresponds to a
spatially-flat universe with with an initially scale-invariant
spectrum and a cosmological constant $\lambda$. Free parameters in the joint
model are the mass density due to all matter ($\Omega$), Hubble's
parameter ($h = H_0 / 100$ km/sec), IRAS light-to-mass bias
($b_{iras}$) and the variance in the mass density field measured in an
$8 h^{-1}$ Mpc radius sphere ($\sigma_{8}$).  For fixed baryon density
$\Omega_b = 0.0125/h^2$ the joint optimum lies at $\Omega= 1 - \lambda
= 0.41\pm{0.08}$, $h = 0.46\pm{0.06}$, $\sigma_8 = 0.65\pm{0.10}$,
$b_{iras} = 1.26\pm{0.06}$ (marginalised error bars correspond to 95
percent confidence).  For these values of $\Omega, \lambda$ and $H_0$
the age of the universe is $\sim 18.7$ Gyr.

 \section {Discussion}

 We have shown some recent studies 
 of  galaxy  surveys, and their cosmological implications. 
New measurements of galaxy clustering and background
radiations can provide improved constraints on the isotropy and
homogeneity of the Universe on large scales.  In
particular, the angular distribution of radio sources and the X-Ray
Background probe density fluctuations on scales intermediate between
those explored by galaxy surveys and CMB 
experiments.  On scales larger than $300 h^{-1} $ Mpc the distribution of
both mass and luminous sources satisfies well the `Cosmological
Principle' of isotropy and homogeneity.  Cosmological parameters 
such as $\Omega$ therefore have a well defined meaning.
 With the dramatic increase of data, we should soon be able to map
the fluctuations with scale and epoch, and to analyze jointly LSS and
CMB data.

\bigskip

 { \bf Acknowledgments} 
 I thank my collaborators for their contribution to the work
 presented here. I also thank 
 the conference organizers for the stimulating and enjoyable meeting, 
 and acknowledge the hospitality of Tokyo University, 
 where this paper was written. 


%


\begin{thebibliography}
%
%

\bibitem{}{}{}
Baleisis, A., Lahav, O., Loan, A.J., Wall, J.V. (1998): 
MN, in press, astro-ph/9709205

\bibitem{}{}{}
Baugh C.M., Efstathiou G., (1994): MN, {\bf 267}, 323

\bibitem{}{}{}
Boldt, E. A. (1987): Phys. Reports, {\bf 146}, 215

\bibitem{}{}{}
Cress C.M., Helfand D.J., Becker R.H., Gregg. M.D., White, R.L.
(1996): {\bf 473}, 7 

\bibitem{}{}{}
Dekel, A., (1994): ARAA, {\bf 32}, 371

\bibitem{}{}{}
Fabian, A. C., Barcons, X. (1992): ARAA, {\bf 30}, 429

\bibitem{}{}{}
Folkes, S., Lahav, O.,  Maddox, S.J. (1996): MN, {\bf 283}, 651

\bibitem{}{}{}
Gawiser, E., Silk, J. (1998): submitted to Science

\bibitem{}{}{}
Gunn, J.E., Weinberg, D.H. (1995): 
           in {\it Wide-Field Spectroscopy and the Distant Universe}, 
           eds. S.J. Maddox \& A. Aragon-Salamanca, World Scientific, 

\bibitem{}{}{}
Hamilton , A. J. S. (1997): 
review to appear the 
{\it Ringberg Workshop on Large-Scale Structure}, 
in Hamilton, D. (ed.), Kluwer Academic, Dordrecht,  astro-ph/9708102 



\bibitem{}{}{}
Kaiser N. (1984): ApJ, {\bf 284}, L9

\bibitem{}{}{}
Kochanek, C.S. (1996): ApJ, {\bf 466}, 638

\bibitem{}{}{}
Lahav O., Piran T., Treyer M.A. (1997): MN, {\bf 284}, 499

\bibitem{}{}{}
Magliocchetti, M.,  Maddox, S.J., Lahav, O.,   Wall, J.V. (1998): 
submitted to MN, astro-ph/9802269

\bibitem{}{}{}
Perlmutter, S., et al.  (1998): Nature, {\bf 391}, 51 

\bibitem{}{}{}
Strauss M.A., Willick J.A.  (1995): Phys. Rep.,
 {\bf 261}, 271

\bibitem{}{}{}
Treyer, M., Scharf, C., Lahav, O., 
Jahoda, K.,  Boldt, E., Piran, T. (1998): 
submitted to ApJ, astro-ph/9801293

\bibitem{}{}{}
Webster, A.M., Lahav, O., Fisher, K.B. (1997): MNRAS, {\bf 287}, 425

\bibitem{}{}{}
Webster, M., Hobson, M.P., Lasenby, A.N., 
Lahav, O.,  Rocha, G., Bridle, S. (1998):
submitted to ApJ Lett, astro-ph/9802109

\bibitem{}{}{}
White, S.D.M., Navarro, J.F., Evrard, A.E., Frenk, C.S. (1993):
{\bf 366}, 429

\bibitem{}{}{}
Wu, K.,K.S.,  Lahav, O.,  Rees, M.J. (1998): submitted to Nature, 
astro-ph/9804062


\end{thebibliography}
\end{document}